\documentclass[%
 reprint,
superscriptaddress,
 amsmath,amssymb,
 aps,
pra,
]{revtex4-2}

\usepackage{graphicx}
\usepackage{dcolumn}
\usepackage{bm}
\usepackage{hyperref}
\usepackage{siunitx}
\usepackage{physics}
\usepackage[version=4]{mhchem}
\usepackage{natbib}
\usepackage[caption=false]{subfig}

\DeclareSIUnit\year{yr}

\begin{document}

\preprint{APS/123-QED}

\title{Laser Cooling of Transition Metal Atoms}

\author{Scott Eustice}
 \email{scott_eustice@berkeley.edu} \affiliation{Department of Physics, University of California, Berkeley, CA 94720}
 \affiliation{Challenge Institute for Quantum Computation, University of California, Berkeley, CA  94720}
\author{Kayleigh Cassella}%
 \affiliation{Department of Physics, University of California, Berkeley, CA 94720}
 \affiliation{Atom Computing Inc., Berkeley, CA 94710}
\author{Dan Stamper-Kurn}
 \email{dmsk@berkeley.edu} \affiliation{Department of Physics, University of California, Berkeley, CA 94720}
 \affiliation{Challenge Institute for Quantum Computation, University of California, Berkeley, CA  94720}
 \affiliation{Materials Science Division, Lawrence Berkeley National Laboratory, Berkeley, CA 94720}

\date{August 17, 2020}

\begin{abstract}
We propose the application of laser cooling to a number of transition-metal atoms,  allowing numerous bosonic and fermionic atomic gases to be cooled to ultra-low temperatures.  The non-zero electron orbital angular momentum of these atoms implies that strongly atom-state-dependent light-atom interactions occur even for light that is far-detuned from atomic transitions.  At the same time, many transition-metal atoms have small magnetic dipole moments in their low-energy states, reducing the rate of dipolar-relaxation collisions.  Altogether, these features provide compelling opportunities for future ultracold-atom research.  Focusing on the case of atomic titanium, we identify the metastable $a  ^5\!F_5$ state as supporting a $J \rightarrow J+1$ optical transition with properties similar to the D2 transition of alkali atoms, and suited for laser cooling.  The high total angular momentum and electron spin of this state suppresses leakage out of the nearly closed optical transition to a branching ratio estimated below $\sim 10^{-5}$.  Following the pattern exemplified by titanium, we identify optical transitions that are suited for laser cooling of elements in the scandium group (Sc, Y, La), the titanium group (Ti, Zr), the vanadium group (V, Nb), the manganese group (Mn, Tc), and the iron group (Fe, Ru).
\end{abstract}

\maketitle


Laser cooling and the achievement of quantum degeneracy of atomic gases has led to an ever broadening range of scientific investigations and applications.  This growing impact on science and technology has been fueled by the availability of quantum gases produced from an increasing number of elements, each of which has a new set of properties that can enable a new family of experiments.  For example, the fortuitous collisional properties of rubidium and sodium enabled the first realizations of scalar \cite{ande95,davi95bec} and spinor \cite{sten98spin,chan05nphys,schm04} atomic Bose-Einstein condensation.  The accessible Feshbach resonances of lithium allowed studies of Efimov states \cite{kram06efimov}.  Isotopes of potassium and lithium allowed the study of resonantly interacting Fermi gases \cite{kina04evidence,rega04res,zwie04pairs,bart04crossover}. The detectability of single metastable helium atoms on micro-channel plate detectors allowed for studies of quantum atom optics \cite{robe01he}.  The narrow lines of alkali-earth atoms and ytterbium enabled the realization of optical lattice clocks \cite{dere11rmp,ludl15rmp}.  The magnetism of chromium allowed for studies of quantum ferrofluids \cite{laha07ferrofluid}, accentuated by the even stronger magnetic dipole interactions of dysprosium \cite{lu11dybec} and erbium \cite{aika12erbium}.  Gaining access to a greater variety of ultracold atomic gases can, therefore, be expected to broaden the impact of ultracold atomic physics even further.

Conversely, the limitations of present-day ultracold atom systems pose limitations on the range of scientific topics that they can be used to study.  As an example, we consider the prospect of studying gases in a stable mixture of internal spin states while subject also to coherent spin-state dependent optical potentials.  Each of these two conditions can be achieved separately in extant quantum gases:  Stable spinor gases are realized with alkali atoms, whose small magnetic moments forestalls inelastic dipolar relaxation collisions \cite{stam13rmp}.  Highly coherent spin-dependent optical potentials are realized for lanthanide atoms, owing to their complex atomic structures \cite{burd16soc}.   However, in neither case are \emph{both} conditions simultaneously achieved:  Spin-dependent optical potentials for alkali atoms have low coherence (as we explain in Sec.\ \ref{sec:titanium}). Spinor gases of lanthanide atoms decay generally through strong magnetic dipolar relaxation, although the specific relaxation via collision channels with indistinguishable initial or final spin states can be suppressed by Fermi statistics \cite{burd15fermionic}.

\begin{figure*}
    \centering
    \includegraphics[width=\textwidth]{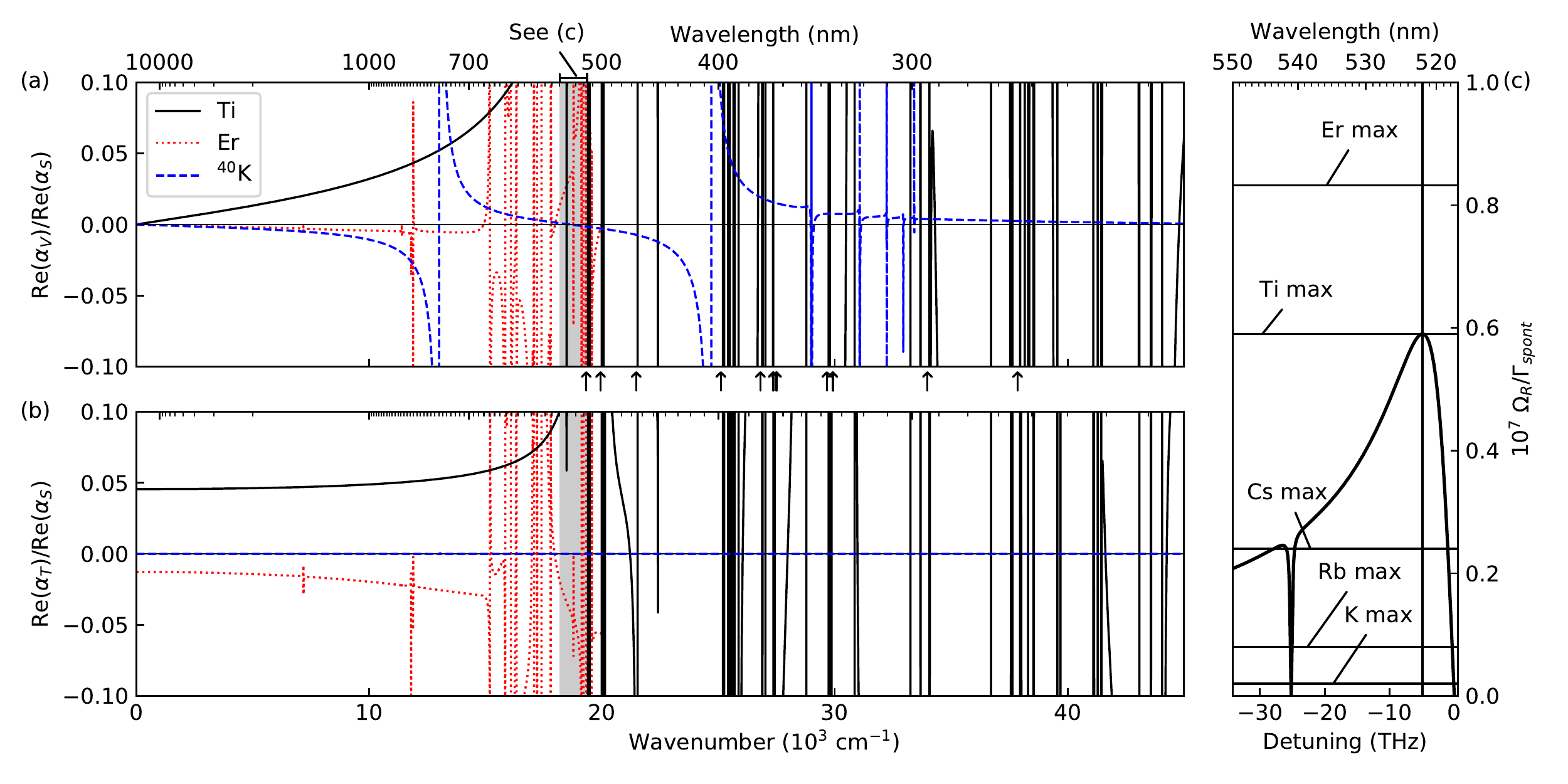}
    \caption{(a) Vector ($\Re \alpha_V$) and (b) tensor polarizability ($\Re \alpha_T$) for Ti (solid black), Er (dashed red), and $^{40}$K (dotted blue) in their absolute ground states, scaled by the scalar polarizability $\Re \alpha_S$ and plotted as a function of the wavenumber and wavelength of applied light. Arrows indicate all transitions to an excited $3d^2\, 4s\, 4p$ state in Ti with an oscillator strength greater than $0.01$. The data for Er, reproduced from Ref.\ \cite{lepe14er}, showing the dense spectrum above 20000 cm$^{-1}$ is omitted for clarity.
    (c) The figure of merit $\mathcal{M} = \Omega_R/\Gamma_\mathrm{sp}$ is plotted over the narrow wavelength range indicated by a gray band in (a, b), with detuning measured relative to the $a^3F_2\to z^3F^{^\circ}_2$ transition. Solid horizontal lines indicate the maximal $\mathcal{M}$ achievable in the Raman coupling scheme for each element, while the vertical line indicates the corresponding wavelength ($\sim$ 522 nm) in Ti.}
    \label{fig:Ti_ac_stark}
\end{figure*}

Here, we open a door to new studies of quantum atomic gases by describing pathways for laser-cooling a number of transition-metal elements, including those in the scandium group (Sc, Y, La), the titanium group (Ti, Zr, and possibly Hf), the vanadium group (V, Nb), the manganese group (Mn, Tc), and the iron group (Fe, Ru, and possibly Os).  Specifically, we find for all these elements that there is a strong, electric-dipole allowed, optical transition (see Tab.\ \ref{tab:alltransitions}), with linewidth on the order of 10 MHz, which resembles the D2 line of alkali atoms in that an electron is driven from $n s_{1/2}$  to the $n p_{3/2}$ state. The lower level on this transition, which we call the laser-cooling state, is either the atomic ground state (in Ru and Mn) or a metastable excited state (in the other cases).  In all cases, these transitions are cycling, or at least very nearly so, and connect states with total angular momentum $J \rightarrow J+1$.  As such, these transitions are suitable for standard laser cooling techniques such as Zeeman slowing \cite{phil82}, magneto-optical trapping \cite{raab87}, and polarization-gradient cooling \cite{dali89pol,lett89}.

These elements have atomic properties that differ from those of existing ultracold atomic gases.  Present-day quantum gases can be divided into two categories: spherical atoms, and highly non-spherical atoms.  Specifically, in the spherical-atom category we include alkali atoms, alkali-earth atoms, metastable noble gases, and additional filled-shell (Yb) or half-filled-shell (Cr) atoms, for which the total orbital electronic angular momentum in the ground state is $L=0$.  In the category of highly non-spherical atoms we identify the lanthanide atoms that have been laser cooled \cite{lu10dytrap,mccl06erbium,miao14ho} and brought to quantum degeneracy \cite{lu11dybec,aika12erbium}.  For these atoms, the ground-level electronic angular momentum is large, e.g., $L=6$ for Dy and $L=5$ for Er.

The transition-metal elements identified in this work have properties that are intermediate to these two categories. For all these elements, both the laser-cooling state and, with the exception of Mn and Tc, also the ground state have $L\neq 0$.  As we discuss below, as for the lanthanide atoms, this condition leads to highly coherent anisotropic light-atom interactions with far-off-resonant light.  However, because the orbital angular momentum of these transition-metal elements is smaller than that of lanthanide elements, we expect ultracold collisions to be more similar to those of spherical atoms.  Specifically, we expect ultracold spinor gases formed of many of these atoms to be stable against inelastic dipolar relaxation collisions.  Altogether, following previous examples, we expect the addition of this new family of transition-metal elements to the rank of ultracold atomic gases will enable novel developments in ultracold atomic (and also molecular) physics.

This paper is organized as follows.  In Section \ref{sec:titanium}, we illustrate our proposed laser cooling scheme by focusing deeply on the example of atomic titanium.  Through this example, we explain how the laser cooling level is identified as a highly spin-polarized configuration with a single electron in the valence s orbital.  The case of titanium also exemplifies experimental techniques required to work with ultracold transition-metal atoms, and points to several potential scientific uses for quantum-degenerate transition-metal atomic gases. In Section \ref{sec:a5f5}, we describe laser cooling pathways for other atoms in the titanium group and in the iron group, which are all similar to the case of titanium.  These include atomic iron, for which laser cooling has already been demonstrated \cite{crau16iron} by the approach described in this paper.  We conclude in Section \ref{sec:otheratoms} by describing briefly the opportunities for laser cooling atoms in the scandium, vanadium and manganese groups.

\section{Titanium and its comparison to existing quantum gases}
\label{sec:titanium}

Atomic titanium (Ti) has four valence electrons surrounding a closed-shell core.  In its lowest-energy term $a ^3F$, Ti has an ``alkali-earth-like'' configuration $3 d^2 \, 4 s^2$. This level is split by fine structure into states with total angular momentum $J=2$, $3$ and $4$, with the $a ^3F_2$ state being the overall ground state.

We outline several reasons why atomic titanium is an attractive choice for ultracold-atom experiments.  

\subsection{Anisotropic optical polarizability}

The non-zero ($L=3$) orbital angular momentum of the ground state of Ti indicates that light-atom interactions remain anisotropic even for light that is far-detuned from atomic resonances.  We have verified this property by computing the relative strengths $\alpha_S(\omega)$, $\alpha_V(\omega)$, and $\alpha_T(\omega)$ of the scalar, vector, and tensor portions of the ground-state ac Stark shift, respectively, as a function of the optical frequency $\omega$.  Here, we express the total ac Stark shift $\alpha_{m_J}(\omega)$, for an atom in the $m_J$ magnetic sublevel defined with respect to a quantization axis $\mathbf{b}$, as \cite{mana86,dere11rmp}
\begin{multline}
    \alpha_{m_J}(\omega) = \alpha_S(\omega) + \left(\mathbf{k} \cdot \mathbf{b}\right) \mathcal{A}\frac{m_J}{2 J}\, \alpha_V(\omega)
    \\ + \frac{1}{2} \left( 3 |\mathbf{\epsilon} \cdot \mathbf{b} |^2 -1 \right) \frac{3 m_J^2 - J (J+1) }{J (2J-1)} \, \alpha_T(\omega).
\end{multline}
In this expression, $\mathbf{k}$ is the unit wavevector and $\mathcal{A}$ the helicity of the light field, and $\mathbf{\epsilon}$ is the optical polarization vector.

We calculate the ac Stark shift by taking into account all electric dipole optical transitions identified in the NIST Spectral Database \cite{nist19} for neutral titanium, using the reported oscillator strengths for all lines for which such strengths are provided.  In addition, there are several excited fine structure levels for which an oscillator strength is not directly reported, but for which we can estimate such strength based on that reported for a different level of the same configuration and term \cite{condon_theory_1979}. As an example, the $a^3F_2\rightarrow w^3D^{^\circ}_1$ and $a^3F_2\rightarrow w^3D^{^\circ}_2$ transitions have reported line strengths of 0.71 a.u.\ and 0.051 a.u.\ respectively, while the line strength of the $a^3F_2\rightarrow w^3D^{^\circ}_3$ transition is not reported. Using the results for LS coupling, we estimate that the unobserved line strength is approximately 0.02 a.u. Our calculation only accounts for transitions with reported or estimated transition rates, and neglects contributions to the atomic polarizability from both the above-ionization continuum and weaker lines without measured transiton rates. Nevertheless, it provides a reasonable estimate for the nature of the ac Stark shifts for long-wavelength ($\lambda \geq 500$  nm) laser light. A similar approach was followed, for example, by Tsyganok et al.\ \cite{tsyg19}

The anisotropy of light-atom interactions is characterized by the ratios $\Re(\alpha_V)/\Re(\alpha_S)$ and $\Re(\alpha_T)/ \Re(\alpha_S)$.  Figure \ref{fig:Ti_ac_stark} shows these calculated ratios for Ti, and compares them to similar ratios calculated for potassium (K) and erbium (Er).  

Examining the long-wavelength limit, i.e.\ far-red-detuned from atomic transitions, we find the polarizability of Ti to be dramatically more anisotropic than for the alkali atom K.  We focus specifically on the isotope $^{40}$K, which, similar to the fermionic isotopes of Ti, is suited to studies of fermionic quantum matter.   In K, two transitions contribute dominantly to the optical response:  the D1 line ($4s ^2S_{1/2}\to4p ^2P^{^\circ}_{1/2}$) at 770.1 nm wavelength and linewidth $2\pi\times6.01$ MHz,  and the D2 line ($4s ^2S_{1/2}\to4p ^2P^{^\circ}_{3/2}$) at 766.7 nm wavelength and linewidth $2\pi\times5.94$ MHz.  Strongly anisotropic polarizability is obtained for light that is near detuned to one or the other of these transitions.  However, for light that is much farther detuned than the $\sim$ THz frequency separation between these fine-structure lines so that the detuning to each of these transitions is nearly the same, the atomic polarizability becomes nearly isotropic.  At a pictorial level, we can consider that the electron distribution in the ground-state of K, with a single valence electron in an s orbital, is isotropic; hence, it responds isotropically to light of any optical polarization.  Coupling between the optical polarization and the electron spin is obtained only by exploiting the spin-orbit coupling (i.e.\ resolving the fine-structure splitting) of the excited state.

In contrast, Ti and also Er show significantly anisotropic interactions in wide parts of the optical spectrum, including far from atomic resonances.  The electronic ground state of Ti is calcium-like, with a strong optical response produced by excitations of the valence $4s^2$ electrons to the $4s \, 4p\, (^1P^{^\circ})$ excited state configuration.  For Ca, this excitation occurs at a single transition at 432 nm wavelength. In contrast, in Ti this excited-state configuration is split into numerous terms (indicated by arrows in Fig.\ \ref{fig:Ti_ac_stark}) because of interactions with the partly filled ($L \neq 0$) $3d$ shell, with the optical coupling strength to each term depending on the optical polarization and the angular momentum state of the $a ^3F_2$ ground-state atom. The excited energy terms are split by large electron-correlation energies, on the order of 100's of THz, leading to large differential detunings for long-wavelength light from each of the relevant excited states.  Altogether, the state-dependent coupling and the large energy differences of these many states results in a strongly anisotropic light-atom interaction.  A similar argument explains the similarly anisotropic polarizability of Er.

An important figure of merit for cold-atom experiments that make use of anisotropic light-atom interactions is the ratio $\mathcal{M} = \Omega_R / \Gamma_\mathrm{sp}$ between the coherent and the incoherent parts of the interaction.  Here, $\Omega_R$ is the Rabi frequency for Raman transitions between different angular-momentum states within the electronic ground state, or, equivalently, the vector or tensor ac Stark shift divided by $\hbar$, while $\Gamma_\mathrm{sp}$ is the total spontaneous emission rate.  Highly coherent (large $\mathcal{M}$) interactions can be achieved for light that is far red-detuned from all optical transitions. Denoting $\mathcal{M}_{V,T}$ for the vector and tensor coupling figures of merit respectively, in the far red-detuned regime, we estimate the average figure of merit as $\mathcal{M}_{V,T}\approx\Re\alpha_{V,T}/|\Im\alpha_S|$.  For example, considering light at a wavelength of 1 $\mu$m, as commonly used in optical-lattice experiments, we expect $\mathcal{M}_V\sim4\times10^5$, $\mathcal{M}_T\sim5\times10^6$ for Ti, compared to $\mathcal{M}_V\sim4\times10^3$, $\mathcal{M}_T\sim 1 \times 10^6$ for Er, and $\mathcal{M}_V\sim1.5\times10^5$, $\mathcal{M}_T\sim1.5$ for $^{40}$K.

Alternately, highly coherent light-atom interactions can be found for light at frequencies that are closer to specific atomic resonances.   For example, Refs.\ \cite{dali16varenna,babi20} propose the use of narrow-line excitations to achieve anisotropic light-atom interactions in Er.   Considering scattering just from the narrow-line resonance itself, the ratio $\mathcal{M}$ generally increases with detuning $\Delta$ between the optical and resonance frequencies, with $\Omega_R \propto \Delta^{-1}$ and $\Gamma_\mathrm{sp} \propto \Delta^{-2}$.  However, at large enough $\Delta$, spontaneous emission from all other atomic resonances becomes dominant, establishing an upper limit of $\mathcal{M}$. Near a particular atomic resonance, the figure of merit can be approximated as \cite{dali16varenna, babi20}
\begin{equation}
    \mathcal{M}\approx\frac{2\Delta_i}{\Gamma_i + \sum_{j\neq i}\Gamma_j\frac{\Delta_i^2}{\Delta_j^2}\frac{A_j}{A_i}}
\end{equation}
where $\Gamma_i$ is the optical detuning from the particular transition chosen, and $\Gamma_j$ is summed over all other transitions.  For Er,  this expression gives a maximum of $\mathcal{M} \sim 8\times10^6$. A similar approach appears possible in atomic Ti.  For example, examining the spectrum of Ti, we find a similar figure of merit, $\mathcal{M}\sim 6 \times 10^6$ for generating state-dependent optical interactions with light that is detuned from the transition between the atomic ground state and the $3d^2(^3F)4s4p(^3P^{^\circ}) z^3F_2^{^\circ}$ excited state, with wavelength $518$ nm.  It is notable that this highly coherent, anisotropic light-atom interaction is achieved with light near the 530 nm wavelength range that is easily produced at high power by frequency-doubled laser systems.  

\subsection{Magnetic dipolar relaxation}

While Ti and the lanthanides Er and Dy both allow for highly coherent anisotropic light-atom interactions, Ti, in its $a ^3F_2$ ground state, has the additional advantage of being relatively stable against inelastic magnetic-dipolar relaxation.  The rate of dipolar relaxation induced by the long-range dipole-dipole interaction in ultracold gases, as considered in Refs.\ \cite{kaga81,shly94,hans03relaxation}, scales (for the magnetic dipole - magnetic dipole case) as the square of the magnetic moment $\mu = g_J J \mu_B$, with $g_J$ being the Land\'{e} g-factor, $J$ being the atomic total angular momentum, and $\mu_B$ being the Bohr magneton.  Magnetic dipolar relaxation in highly magnetic atoms such as Dy \cite{burd15fermionic} and Cr \cite{hans03relaxation}, with magnetic moments of $\mu = 10$ and $6 \, \mu_B$, respectively, is rapid, leading to spin-polarization lifetimes on the order of 1 second or below under common experimental gas densities ($\sim10^{13} \, \mbox{cm}^{-3}$) and applied magnetic fields ($\sim1$ G).  In contrast, for alkali atoms, with magnetic moments of $1 \mu_B$ and below, the minutes-long stability of magnetically trapped atoms indicates the dipolar relaxation rate to be very small.

For Ti, the $J=2$, $3$, and $4$ ground states have magnetic moments of $g_J J \mu_B = 4/3$, $13/4$, and $5 \, \mu_B$, respectively.  Since the magnetic moment of the overall ($J=2$) ground state is similar in magnitude to that of an alkali atom, we expect dipolar relaxation in a $a ^3F_2$ Ti gas to be very slow.  Additional inelastic decay paths could be opened by interactions between the open d shells of the colliding atoms.  However, several studies, e.g.\ theoretical studies on Sc \cite{karm14scandium1,karm14scandium2} and experiments on buffer-gas-cooled Ti gases \cite{lu09ti}, indicate that such inelastic processes are suppressed by the fact that the 3d shell is submerged below the closed 4s shell.  Altogether, collisions among ground-state Ti atoms should be predominantly spin-preserving, allowing, for example, for magnetic trapping in weak-field seeking states and for studies of spinor gases with conserved magnetization.

\subsection{Isotopic richness}

Another favorable property of Ti is that it is endowed with numerous stable isotopes, presenting a broad palette of choices for ultracold atomic gas experiments.  These choices include three bosonic isotopes, $^{46, 48, 50}$Ti, with relatively large natural abundance (8\%, 74\% and 5\%, respectively).  All three bosonic isotopes have nuclear spin $I=0$, leading to a simple atomic level structure that is free of hyperfine structure.  As quantum degenerate gases, these three isotopes, in their $a ^3F_2$ ground state, may present examples of $J=2$ spinor Bose-Einstein condensates, which are predicted to manifest three different types of magnetic ordering: ferromagnetic, nematic and tetragonal \cite{ciob00,koas00,ueda02spin2}.  The choice of ground state depends on the relative strengths of three s-wave scattering lengths that characterize the low-energy collisions between $J=2$ bosons.  Both the nematic and tetragonal phases, owing to their discrete rotational symmetries, support structures such as fractional superfluid vortices \cite{make03defects,seme07onethird,song07}, whose non-abelian character can lead to complex vortex collision and rung dynamics \cite{koba09nonabelian}.  While $^{87}$Rb does already permit studies of $F=2$ spinor gases in its higher-energy hyperfine spin manifold, such gases have only limited lifetimes (of order 100 ms at typical densities) owing to hyperfine relaxation collisions.

Ti also has two stable fermionic isotopes, $^{47,49}$Ti, with nuclear spins $I=5/2$ and $7/2$, and abundances of $7\%$ and $5\%$, respectively.  Given the strong anisotropy of light-atom interactions, and the stability of spin mixtures against dipolar relaxation, these isotopes could be used to study Cooper pairing in spin-orbit coupled Fermi gases. Such systems, which are under investigation also within solid-state devices, are predicted to support topological superfluidity and Majorana fermions \cite{reed00,ivan01,fu08}, with relevance to topological quantum computing \cite{kita03,ster08review,naya08rmp}.

\subsection{Laser cooling scheme}

\begin{figure*}
    \centering
    \includegraphics[width=0.2365\textwidth]{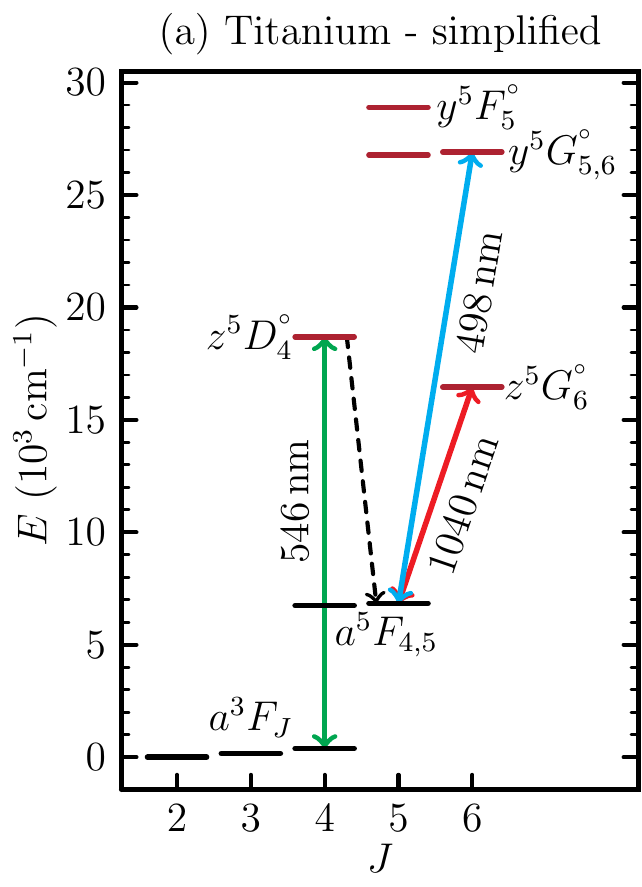}\hspace{0.002\textwidth}\includegraphics[width=0.20615\textwidth]{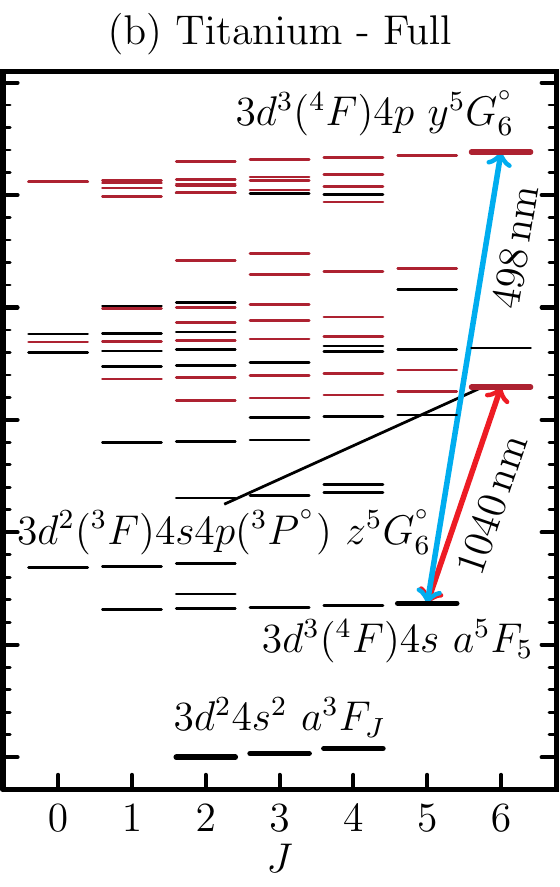}\hspace{0.002\textwidth}\includegraphics[width=0.20615\textwidth]{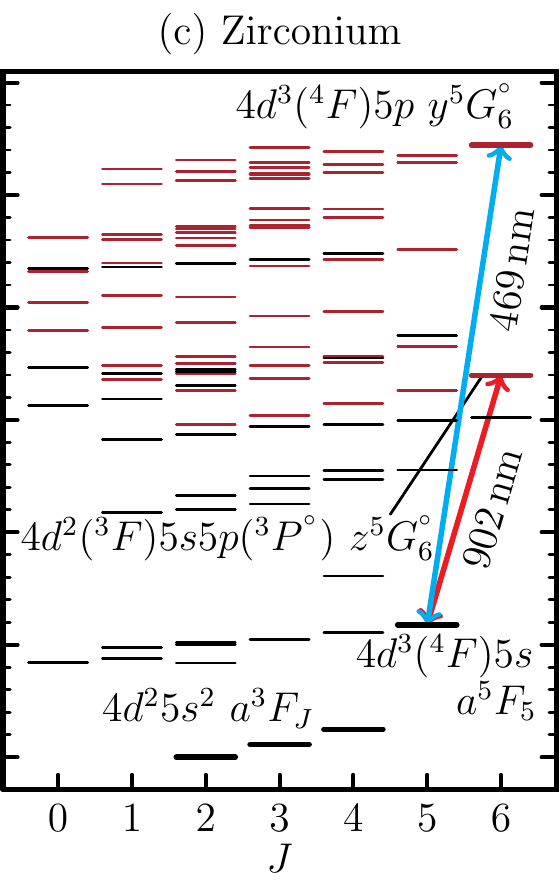}
    \caption{Energy levels of titanium group elements Ti and Zr. Here, and in subsequent energy-level figures, black (red) lines indicate even (odd) parity states, and thicker lines highlight states relevant for laser cooling. Levels are arranged by total electronic angular momentum $J$ and ordered by energy. (a) Shows a simplified energy level diagram of Ti, highlighting the important levels and transitions involved in the entire laser cooling scheme. The $a^3F_J$ fine structure ground levels and $a^5F_5$ metastable laser-cooling level are the relevant even parity states. The $z^5D^{^\circ}_4$ level is one of several intermediate states that could be used to optically pump atoms into the $a^5F_5$ state, with the 546 nm transition indicated as well as the dominant decay path. The two excited levels for laser cooling, the $z^5G^{^\circ}_6$ and $y^5G^{^\circ}_6$, are also identified and can support type I MOTs with transitions at 1040 nm (red) and 498 nm (blue) respectively. (b) A more complete diagram of Ti levels is given, including the full configuration and term symbol for the ground states and laser cooling excited states. For simplicity, optical pumping transitions are suppressed. All levels below the $y^5G_6^{^\circ}$ state are shown. (c) Zr level structure.  All levels below the $y^5G^{^\circ}_6$ are shown.  The laser-cooling transitions, analogous to those in Ti, are shown as well.}
    \label{fig:Ti_group_levels}
\end{figure*}

Many successful realizations of laser cooling make use of atomic transitions that are approximately closed, driving an atom from a low-energy state of angular momentum $J$ to an excited state of angular momentum $J+1$.  The closed nature of the transition ensures than an atom will scatter the large number of photons ($10^3$ or greater) required to reduce its velocity through the optical scattering force, returning always to the laser-cooling state.  The $J \rightarrow J+1$ nature of the transition provides for optical pumping of the atom into a brightly scattering state and enables techniques such as magneto-optical trapping and polarization-gradient cooling.

There is no $J\rightarrow J+1$ transition from the $a ^3F$ ground term of titanium that satisfies these criteria.  The optical spectrum of this ground term, dominated by excitation of one of the $4 s^2$ electrons of the ground configuration, resembles that of calcium.  As in the case of calcium \cite{kuro90casr,kist94calcium,oate99ca}, laser cooling via one of these transitions is unfavorable since there is significant leakage from the excited states to other even-parity metastable states.

Instead, a suitable laser-cooling transition is found among the family of lines that arise after exciting an electron to a metastable alkali-like configuration $3 d^3  4 s^1$. The lowest energy level of this metastable configuration, the $a^5F$ level, is split via fine structure into $J=1$ to $J=5$ states. While the lifetime of these states has not been measured, the two dominant decay mechanics, magnetic dipole radiation between fine structure states and spin-forbidden electric quadrupole decay to the ground state, are weak transitions. Comparing to similar atoms, such as Fe, we expect the radiative lifetime of these states to be on the order of 10-100 s \cite{nist19}.  The highest-$J$ state, labeled $a ^5F_5$ and with an energy of 6843 cm$^{-1}$ above the ground state, has a strong $J \rightarrow J+1$ transition to the odd-parity $y ^5G_6^{^\circ}$ state (at energy 26911 cm$^{-1}$).  On this transition, at a wavelength of 498.3 nm, the valence $4s_{1/2}$ electron is excited to the $4p_{3/2}$ level. The remaining valence electrons remain in the spin polarized $3 d^3 (^4F_{9/2})$ configuration.  As such, the transition resembles the D2 $4s \rightarrow 4p$ transition in potassium if we regard Ti's $3 d^3$ valence electrons as part of the inert atomic core.  The linewidth of this transition ($\Gamma = 2\pi\times\SI{10.5}{\mega\hertz}$) is comparable to alkali atom cooling transitions, with a corresponding Doppler cooling limit of $T_{\text{D}}= \hbar \Gamma / (2 k_B) = \SI{250}{\micro\kelvin}$. This line is amenable to all the standard tricks of laser-cooling: Zeeman slowing, magneto-optical trapping, and polarization-gradient cooling. Furthermore, the line is sufficiently broad to provide a strong cooling force.

This transition appears to be a cycling transition because of the spin-polarized nature of the $3d^3$ electrons and the maximal $J=5$ value of the total electronic angular momentum.  Electric dipole transitions could, in principle, lead from the odd-parity excited state to four other even-parity states: the $a ^3G_5$ state (at energy of 15220 cm$^{-1}$), the $a ^3H_{5,6}$ states (18141 cm$^{-1}$ and 18192 cm$^{-1}$, respectively), and the $a^1H_5$ state (20796 cm$^{-1}$).  However, these transitions are all spin forbidden, requiring a reconfiguration of the $3d^3$ valence electrons. 
Focusing on the leakage into states with $J=5$, we estimate the residual branching ratio for spontaneous emission into these states by considering the degree to which a spin-orbit interaction term would mix these states with the $a ^5F_5$ state.  We quantify the strength of the spin-orbit mixing energy from the fine-structure splitting of the $a^5F_5$ level, and then apply second-order perturbation theory to estimate the branching ratio as $P_i=\delta^2/\Delta_i^2$, where $\delta=E_{a^5F_5}-E_{a^5F_4}$ is the fine structure splitting and $\Delta_i=E_{i}-E_{a^5F_5}$ is the difference in energy between a given leakage state $i$ and the $a^5F_5$ state. Under these assumptions, we expect that the total leakage $P=\sum_iP_i=\sum_i\frac{\delta^2}{\Delta_i^2}$ from the cooling transition to be around $3 \times 10^{-5}$, low enough to perform laser cooling and trapping without the need for constant repumping lasers. For the remaining atoms being considered, the leakage is also small and not reported in any spectroscopic data. We calculated the leakage through this same method of estimating the mixing between states of different spins by the ratio $\delta/\Delta_i$ for the given possible leakage states.

The $a^5F_5$ state supports an additional cycling transition.  This transition to the $z ^5G_6^{^\circ}$ state has an optical wavelength of $1040$ nm and a linewidth of $2\pi\times\SI{11.3}{\kilo\hertz}$, giving a Doppler temperature of $T_{\text{D}}=\SI{0.27}{\micro\kelvin}$. As has been demonstrated with Sr \cite{kato99,voge99}, Ca \cite{binn01}, Yb \cite{kuwa99}, Dy \cite{lu11dybec}, and Er \cite{fris12narrow}, a second-stage magneto-optical trap using this narrower line could significantly increase the phase space density of the laser cooled gas.  Furthermore, there are even fewer leakage channels for emission out of the $z ^5G_6^{^\circ}$ state than out of the $y ^5G_6^{^\circ}$ state, increasing the number of photons that could be scattered without the need for repumping. By the same method of estimation, we expect the leakage from this transition to be around $1\times10^{-7}$.

Producing an atomic beam of Ti through direct sublimation requires a Ti source to be heated to temperatures upwards of 1500 K.  At this high temperature, almost $0.1\%$ of the sublimated atoms will be thermally excited into the metastable laser-cooling state. A higher population in the laser-cooling state can be achieved by optical pumping of ground-level atoms. To find a promising excited state $e$ for efficient optical pumping from a ground state $g$ to a target state $t$, we have examined the available data on the NIST database. Under the assumptions that we have sufficient laser power to drive the optical pumping transition and that the leakage out of the optical pumping transition to other metastable dark states $d_i$ is sufficiently captured by existing data, we calculate the efficiency of an optical pumping transitions as
\begin{equation}
    \eta=\frac{A_{eg}+A_{et}}{A_{eg}+A_{et}+\sum_iA_{ed_i}}=\frac{A_{eg}+A_{et}}{\Gamma_e}
\end{equation}
In titanium, we identify three promising schemes for transferring atoms from the $a ^3F_4$ state into the $a ^5F_5$ state: one-color optical pumping via the $z ^5D_4^{^\circ}$ state (wavelength $\lambda = 546.20$ nm, linewidth $\Gamma = 2\pi \times 0.3$ MHz, transfer efficiency $\eta = 0.88$), one-color optical pumping via the $y ^5F_5^{^\circ}$ state ($\lambda = 350.76$ nm, $\Gamma = 2 \pi \times 16$ MHz, $\eta = 0.9$), or two-color pumping via the $y ^5G_5^{^\circ}$ state ($\lambda=378.99$ nm, $\Gamma=2\pi\times12.7$ MHz) and $y ^5F_5^{^\circ}$ state ($\lambda=451.40$ nm, $\Gamma=2\pi\times15.7$ MHz).  In the two-color scheme, the first transition pumps the majority of the atoms into the $a ^5F_4$ state, which is then pumped by the second transition to the laser coolable $a ^5F_5$ state, for a total efficiency of $\eta = 0.995$. There are multiple routes to transfer atoms from the $a^5F_5$ state to the $a^3F_2$ ground state; one simple path would be to drive the $a^5F_5\rightarrow z^3F_4^{^\circ}$ transition at 758 nm, which transfers atoms to the $a^3F_4$ and $a^3F_3$ states. Simultaneously driving the $a^3F_4\rightarrow y^3F_3^{^\circ}$ and $a^3F_3\rightarrow y^3F_2^{^\circ}$ transitions at the wavelengths of 403 nm and 401 nm leads to a total transfer efficiency of $\eta = 0.814$ for transferring $a^5F_5$ atoms to the $a^3F_2$ ground state. More efficient schemes come at the cost of greater experimental complexity.

\section{Titanium and iron group atoms}
\label{sec:a5f5}

\subsection{Titanium group}

Other elements in the titanium group, zirconium (Zr) and hafnium (Hf), have similar atomic structures and spectra to Ti. There are two low-energy, even-parity electronic configurations, in which the four valence electrons reside in either an alkali-earth-like configuration of $(n-1)d^2\, ns^2$ or an alkali-like configuration $(n-1)d^3\, ns^1$ where $n$ is the principal quantum number of the valence s shell. The ground level has the configuration $(n-1)d^2\, ns^2$ and term symbol $a^3F_2$, with fine-structure splitting to other $J$ levels on the order of 100 cm$^{-1}$. The lowest-energy state with the $(n-1)d^3\, ns^1$ electron configuration has a term symbol $a^5F$ and the spin-polarized laser-cooling state is identified as  the highest fine-structure state with $J=5$. 

\paragraph{Zirconium:}

The level structure of Zr is shown in Fig.\ \ref{fig:Ti_group_levels}.  The laser-cooling state lies 5889 cm$^{-1}$ above the ground state.  A strong laser-cooling transition to the $y ^5G_6^{^\circ}$ state lies at a wavelength of $468.9$ nm.  An additional closed cooling transition, to the $z ^5G_6^{^\circ}$ state, occurs at the wavelength $901.8$ nm.  Line widths for these two transitions have not been measured.  However, one expects linewidths similar to those found in Ti, i.e.\ that the blue-light transition, which involves a $(n-1)d^3ns \rightarrow (n-1)d^3\, np$ electron configuration change, will have a linewidth in the range of $2 \pi \times 10$ MHz, while the red transition, which involves a $(n-1)d^3\, ns \rightarrow (n-1)d^2\, ns\, np$ core-electron configuration change, will be substantially narrower than blue-light transition. Given the increased spin-orbit coupling of Zr, leakages from the excited states are expected be greater than in Ti. We expect the leakage for the broad (narrow) transition to be around $1\times10^{-4}$ $(1\times10^{-5})$. Zr possesses six stable or long-lived naturally abundant isotopes: four bosonic isotopes ($^{90,92,94,96}$Zr) with nuclear spin $I=0$, and two fermionic isotope ($^{91,93}$Zr) with $I=5/2$.

\paragraph{Hafnium:}
The spectrum of Hf is too poorly known to determine whether the analogous laser-cooling transition exists below the first ionization limit (55047.9 cm$^{-1}$). However, there is a metastable state that should be long lived and might support laser cooling: the $5d^3(b\;^4F)6s\;a^5F_5$ state with an energy of 17901.28 cm$^{-1}$ above the ground state.

\subsection{\label{sec:lasercoolingFe}Iron group}
The elements in the iron group, iron (Fe), ruthenium (Ru), and osmium (Os), also have two low energy even parity electronic configurations: an alkali-earth-like valence configuration $(n-1)d^6\, ns^2$ and the alkali-like valence configuration $(n-1)d^7\, ns^1$. As in the titanium group, a cycling transition exists on the maximum-$J$ level of the lowest-energy alkali-like level, also given by the term symbol $a^5F_5$. However, because the valence d-orbital of the iron group is more than half filled, metastable states do not enjoy the protection from decay afforded by a required spin-flip; the total spin $S$ of the ground and metastable cooling states are equal. 

\paragraph{Iron:}
\begin{figure}
    \centering
    \includegraphics[width=0.25\textwidth]{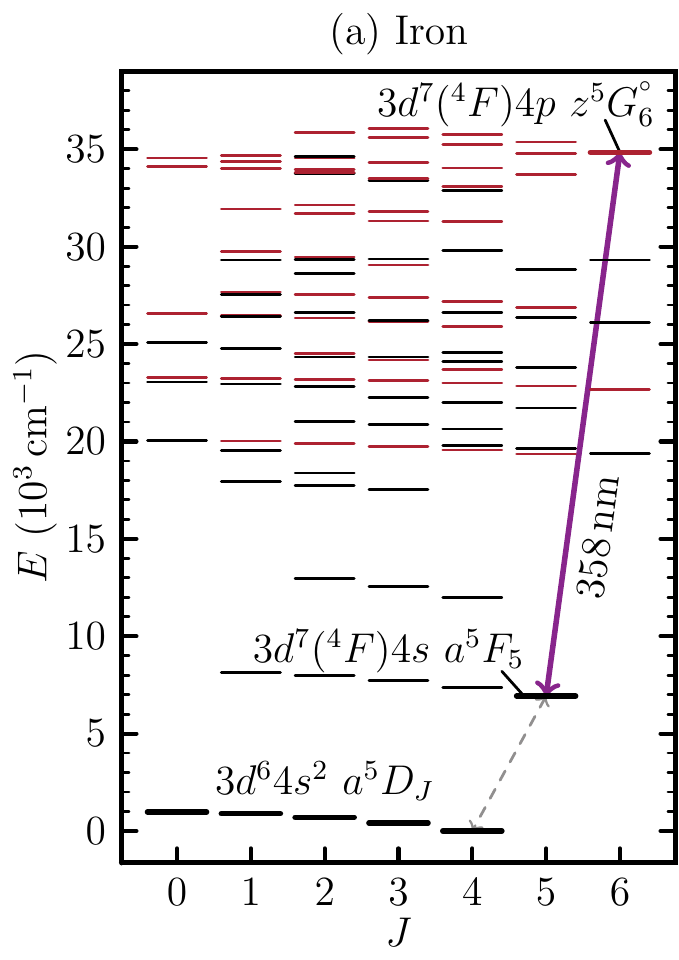}\hspace{0.002\textwidth}\includegraphics[width=0.20615\textwidth]{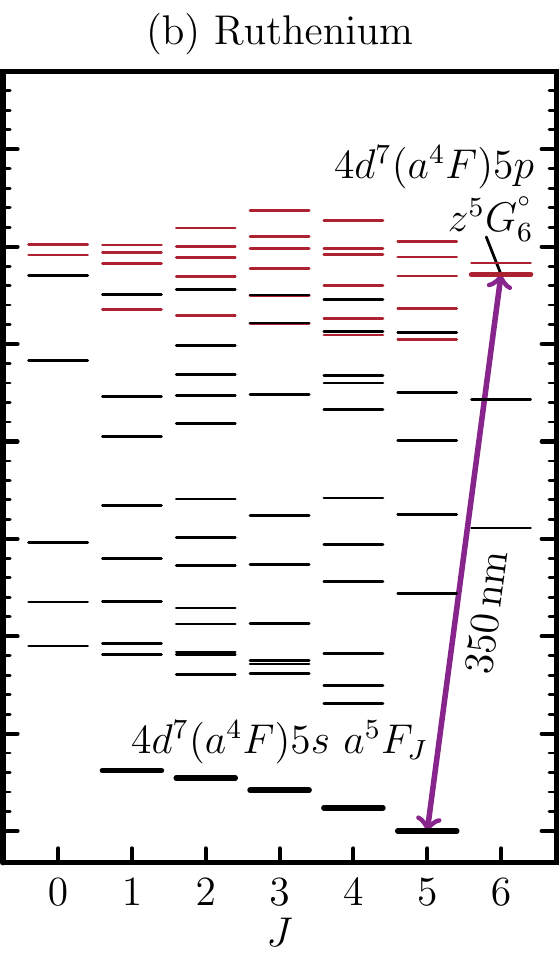}
    \caption{Energy levels of iron group elements (a) Fe and (b) Ru. In Fe, the metastable laser-cooling state is the $a ^5F_5$ state. The laser-cooling transition at 358 nm wavelength drives the atom to the $z ^5G_6^{^\circ}$ excited state.  Laser cooling on this transition has been demonstrated \cite{huet15iron,crau16iron}. Indicated by the dashed gray arrow is the decay of the $a^5F_5$ state in Fe, predicted to occur at a rate of $2 \times 10^{-3} \, \mbox{s}^{-1}$ \cite{nist19}. In Ru, the $a^5F_5$ laser-cooling state is also the overall ground state, eliminating the need for optical pumping into a metastable state.  The 350 nm wavelength laser-cooling transition, connected to the $z^5G^{^\circ}_6$ state is indicated.}
    \label{fig:Fe_group_levels}
\end{figure}

The laser cooling of iron (Fe) has been realized by the Bastin group \cite{huet15iron}. The ground level of Fe has an electron configuration of $3d^6\, 4s^2$ and term symbol $a^5 D_J$, with total angular momenta $J$ ranging from $J = 0$ to $J=4$.  The overall ground ground state, $a ^5D_4$, has a large magnetic moment and is expected to undergo significant magnetic dipole-dipole relaxation.  The metastable cooling state $3d^7(^4F)\, 4s$ $a^5F_5$, with an energy of \SI{6928}{\per\centi\meter}, has an estimated lifetime between $300-1000$ s \cite{nist19}, allowing ample time for cooling and trapping of Fe atoms. This state has a closed transition to the excited state $3d^7(^4F)\, 4p$ $z^5G^{^\circ}_6$ with a wavelength of \SI{358}{\nano\meter} and a linewidth of $2\pi\times\SI{16.2}{\mega\hertz}$. 
Similar to Ti, it is possible in Fe to increase the population of atoms in the metastable cooling state via optical pumping; in Ref.\ \cite{crau16iron}, the 372-nm-wavelength transition to the $z ^5F^{^\circ}_5$ state is used.
While not reported in the literature, we expect the leakage from the cooling transition to be around $4\times10^{-5}$.

\paragraph{Ruthenium:}

Ruthenium (Ru) may be a surprisingly convenient choice for laser cooling.  For this atom, the laser-cooling state $a ^5F_5$ is indeed the overall ground state, obviating the need for optical pumping to drive atoms into the laser-cooling  state.  Laser cooling would proceed directly through the transition to the $z^5G_6^{^\circ}$ excited state at a wavelength of \SI{350}{\nano\meter} and with a linewidth of $2\pi\times\SI{13.7}{\mega\hertz}$ . The leakage from this transition is expected to be around $7\times10^{-4}$. 

Ru has several additional features of interest.  First, the spectrum of Ru should contain a narrow-line magnetic dipole transition from the metastable even-parity $4d^7(a^4F)\, 5s$ $a^5F_1$ state up to the low-lying  $4d^6\, 5s^2$ $a^5D_0$ excited state.  This transition frequency should be insensitive at first order  to applied magnetic fields.  Moreover, light at the transition wavelength of 1566 nm could be transmitted at low loss through optical fibers. The excited state, which can decay either on a magnetic dipole transition towards the $a^5F_1$ state, on an electric quadrupole transition to the $a^5F_2$ state, or on a two photon electric dipole transition to the $a^5F_1$ state, should be very long-lived.  Altogether, this transition could offer advantages as an optical clock standard.  Second, Ru has a large number of stable isotopes, including five bosonic isotopes ($^{96, 98, 100, 102, 104}$Ru) with a nuclear spin $I=0$.  The isotope shifts on the various narrow-line transitions in Ru could be leveraged to search for forces outside the Standard Model \cite{bere18}.

\paragraph{Osmium:}

As with Hf, the spectrum of Os is not  known well enough to determine whether there is a closed laser-cooling transition below the first ionization limit. However, the $5d^7(^4F)\, 6s$ $a^5F_5$ metastable state (energy of 5144 cm$^{-1}$) would likely have a long lifetime and could support laser cooling 

\begin{table*}[t]
    \centering
    \begin{ruledtabular}
    \begin{tabular}{ccccccccc}
        \textbf{Atom}&\multicolumn{2}{c}{\textbf{Ground state}}& \multicolumn{2}{c}{\textbf{Metastable state}}&\multicolumn{2}{c}{\textbf{Excited state}}&Cooling&Linewidth\\
        &config.&term&config.&term&config.&term&trans. (\si{\nano\meter})&($\si{\mega\hertz}/2\pi$)\\\hline\\
        Sc&$3d4s^2$&$a^2D_{3/2}$&$3d^2(^3F)4s$&$a^4F_{9/2}$&$3d^2(^3F)4p$&$z^4G_{11/2}^{^\circ}$&567&8.6 \\
        Y&$4d5s^2$&$a^2D_{3/2}$&$4d^2(^3F)5s$&$a^4F_{9/2}$&$4d^2(^3F)5p$ & $z^4G_{11/2}^{^\circ}$&547&10.\\
        La&$5d6s^2$&$^2D_{3/2}$&$5d^2(^3F)6s$&$^4F_{9/2}$&$5d^2(^3F)6p$&$^4G_{11/2}^{^\circ}$&625&5.86\\
        &&&&&$4f5d(^3G^{^\circ})6s$&$^4G_{11/2}^{^\circ}$&406&8.9\\\\
        Ti& $3d^24s^2$ & $a^3F_4$ & $3d^3(^4F)4s$ & $a^5F_5$ & $3d^3(^4F)4p$ & $y^5G_6^{^\circ}$ & 498 & 10.5 \\
        & & & & & $3d^2(^3F)4s4p(^3P^{^{\circ}})$ & $z^5G_6^{^\circ}$ & 1040 & 0.0113\\
        Zr& $4d^25s^2$ & $a^3F_4$ & $4d^3(^4F)5s$ & $a^5F_5$ & $4d^3(^4F)5p$ & $y^5G_6^{^\circ}$ & 469 & \textbf{---}\\
        & & & & & $4d^2(^3F)5s5p(^3P^{^{\circ}})$ & $z^5G_6^{^\circ}$ & 902* & \textbf{---}\\ \\
        V & $3d^34s^2$ & $a^4F_{3/2}$ & $3d^4(^3H)4s$ & $a^4H_{13/2}$ & $3d^4(^3H)4p$ & $z^4I_{15/2}^{^\circ}$ & 445 & 13.\\
        Nb & $4d^4(^5D)5s$ & $a^6D_{1/2}$ & $4d^4(^3H)5s$ & $a^4H_{13/2}$ & $4d^4(^3H)5p$ & $^4I_{15/2}^{^\circ}$ & 413 & \textbf{---} \\
         &  &  &  &  & $4d^35s(^3H)5p$ & $z^4I_{15/2}^{^\circ}$ & 463 & \textbf{---} \\\\
        Mn & $3d^54s^2$ & $a^6S_{5/2}$ & \textbf{---} & \textbf{---} & $3d^5(^6S)4s4p(^3P^{^\circ})$ & $z^6P_{7/2}^{^\circ}$  & 403 & 2.71 \\
        & & & $3d^6(^3H)4s$ & $a^4H_{13/2}$  & $3d^6(^3H)4p$ & $z^4I_{15/2}^{^\circ}$ & 405* & \textbf{---} \\
        Tc & $4d^55s^2$ & $a^6S_{5/2}$ & $4d^6(^3H)5s$ & $^4H_{13/2}$ & $4d^6(^3H)5p$ & $^4I^{^\circ}_{15/2}$ & 378 &\textbf{---}\\ \\
        Fe & $3d^64s^2$ & $a^5D_4$ & $3d^7(^4F)4s$ & $a^5F_5$ & $3d^7(^4F)4p$ & $z^5G_6^{^\circ}$ & 358 & 16.2 \\
        Ru & $4d^7(^4F)5s$ & $a^5F_5$ & \textbf{---} & \textbf{---} & $4d^7(^4F)5p$ & $z^5G_6^{^\circ}$ & 350 & 13.7 \\ \\
    \end{tabular}
    \end{ruledtabular}
    \caption{Laser cooling lines of transition metals. Configurations, terms, and essential cooling line information is summarized. Values with a (*) indicate that the wavelength has been inferred from energy level data, not measured directly. All values are drawn from the NIST database \cite{nist19}.}
    \label{tab:alltransitions}
\end{table*}

\begin{figure*}
    \centering
    \includegraphics[width=0.23\textwidth]{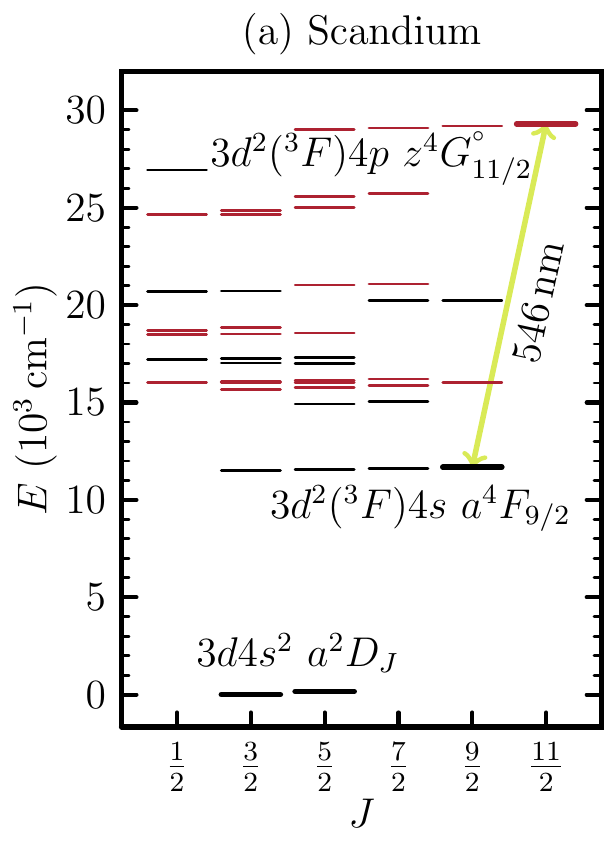}\hspace{0.002\textwidth}\includegraphics[width=0.18475\textwidth]{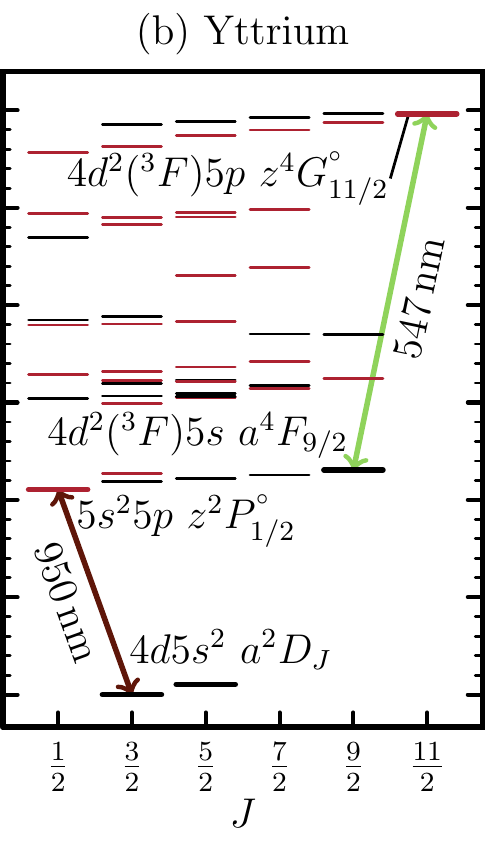}\hspace{0.002\textwidth}\includegraphics[width=0.2129\textwidth]{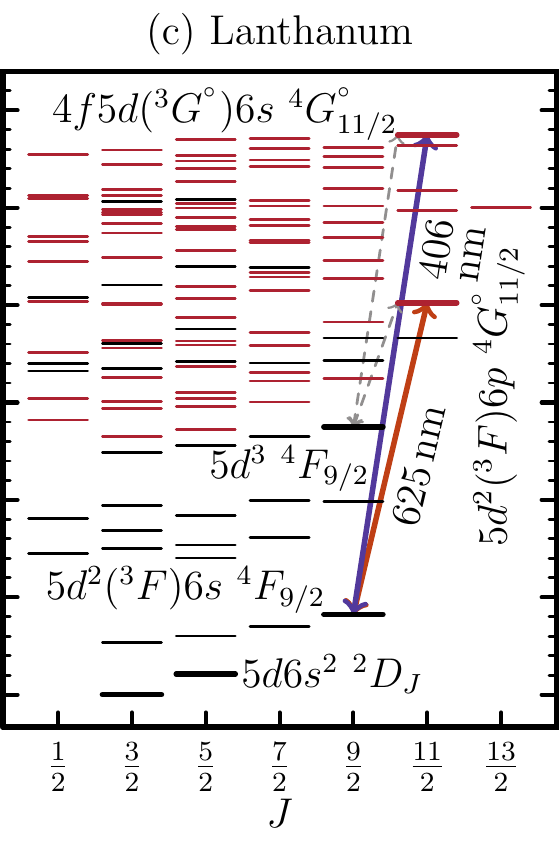}
    \caption{Energy levels of scandium group elements (a) Sc, (b) Y, and (c) La. The $a^2D_J$ ground term and $a^4F_{9/2}$ metastable laser-cooling state are highlighted in all diagrams. For Sc, the laser-cooling transition at 549 nm wavelength connects to the $z^4G_{11/2}$ state. In Y, the analogous transition occurs at wavelength 547 nm; also indicated is the narrow laser-cooling transition for the ground state at wavelength 950 nm. For La, two laser-cooling states at wavelengths 625 nm and 406 nm connect to two $^4G_{11/2}$ states. The grey dashed arrows indicated two potential spin-allowed leakage paths that may require repumping.}
    \label{fig:Sc_group_levels}
\end{figure*}

\section{Scandium, vanadium and manganese group atoms}
\label{sec:otheratoms}

The general method of cooling gases of transition metals proposed herein -- employing a long-lived alkali-like metastable state with configuration $(n-1)d^{x+1}\, ns^1$ consisting of a polarized submerged shell core  and a single valence $s$ electron --  is applicable also to several other transition-metal atoms.  Table~\ref{tab:alltransitions} gives the relevant properties for the metals of the scandium, vanadium and manganese group that we have identified as amenable to laser cooling.  We discuss these cases briefly below.

\subsection{Scandium group}

Atoms in the scandium group -- Sc, Y, and La --  have a ground state characterized by two valence $s$-electrons ($(n-1)d^1\, ns^2$ $a^2D_{3/2}$, where $n$ is the principal quantum number of the valence shell (with $n$ being 4, 5, or 6 for Sc, Y, and La, respectively) and a spin-polarized metastable alkali-like configuration ($(n-1)d^2\, ns^1$ $a^4F_{J}$). The laser-cooling state is given by the highest fine-structure level $J=9/2$ of this metastable state from which a strong, closed $J+1$ transition exists up to an odd parity excited state $z^4G_{11/2}^{^\circ}$ (see Fig.\ \ref{fig:Sc_group_levels} for atomic level structures).

\paragraph{Scandium:} In scandium (Sc), the cooling scheme proceeds without any mitigating factors. The laser-cooling transition occurs at \SI{567.3402}{\nano\meter} \cite{bena77scandium} with a linewidth of $2\pi\times\SI{8.6}{\mega\hertz}$ \cite{lawl89sc}. Sc possesses one stable isotope, bosonic \ce{^{45}Sc}, with nuclear spin $I=7/2$.

\paragraph{Yttrium:} In Y, there is the laser cooling line from a metastable state as described above, with a wavelength of \SI{546.7986}{\nano\meter} \cite{megg75intensities} and a linewidth of $2\pi\times\SI{10.}{\mega\hertz}$ \cite{hann82}. Yttrium possesses one stable isotope, bosonic \ce{^{89}Y}, with nuclear spin $I=1/2$.

A second transition may also be considered, between the overall ground state $4d5s^2$ $a^2D_{3/2}$ and the lowest odd-parity excited state, $5s^25p$ $z^2P_{1/2}^{^\circ}$.  As a $J \rightarrow J-1$ transition, this excitation would be suited to a type II magneto-optical trap \cite{flem97}.  While the transition itself has not yet been observed, its wavelength of 949.7429 nm is known \cite{b._a._palmer_first_1977}.  There are no possible leakage channels for this transition, allowing a clear cycling transition to be used. It would likely be a relatively narrow E1 transition, since it involves the transition of the inner shell $4d$ electron to the $5p$ state.

\paragraph{Lanthanum:} Lanthanum (La) returns to the pattern of only being coolable from a metastable state. However, it has two possible cycling transitions available in this state.  One laser-cooling transition connects the metastable $5d^2(^3F)\, 6s$ $^4F_{9/2}$ and the $5d^2(^3F)\, 6p$ $^4G^{^\circ}_{11/2}$ state, occurring at $\SI{625.166}{\nano\meter}$ \cite{megg75intensities} with a linewidth of $2\pi\times\SI{5.86}{\mega\hertz}$ \cite{hart15la}. A second transition  connects the metastable state to the excited state $4f5d(^3G^{^\circ})\, 6s$ $^4G^{^\circ}_{11/2}$, with wavelength $\SI{406.148}{\nano\meter}$ \cite{megg75intensities} and linewdith $2\pi\times\SI{8.9}{\mega\hertz}$ \cite{hart15la}. Lanthanum possesses one stable isotope, bosonic \ce{^{139}La}, with a nuclear spin $I=7/2$. However fermionic \ce{^{138}La} is long lived, with a half-life of \SI{1.02e11}{\year}, which could allow an isotopically enriched sample of fermionic atoms to be studied. \ce{^{138}La} has a nuclear spin $I=5$.

Lanthanum also has a difficulty not present in the other elements that have been discussed: from both of the excited states that could be addressed for laser cooling, there is a spin-allowed electric dipole transition to the $5d^3$ $^4F_{9/2}$ state. While decay to this state has not been observed, it could be a stronger source of leakage than in the other examples considered thus far. If these leakage channels are significant, they could be repumped to replenish the atoms in the metastable cooling state.

\subsection{Vanadium group}

\begin{figure}
    \centering
    \includegraphics[width=0.26\textwidth]{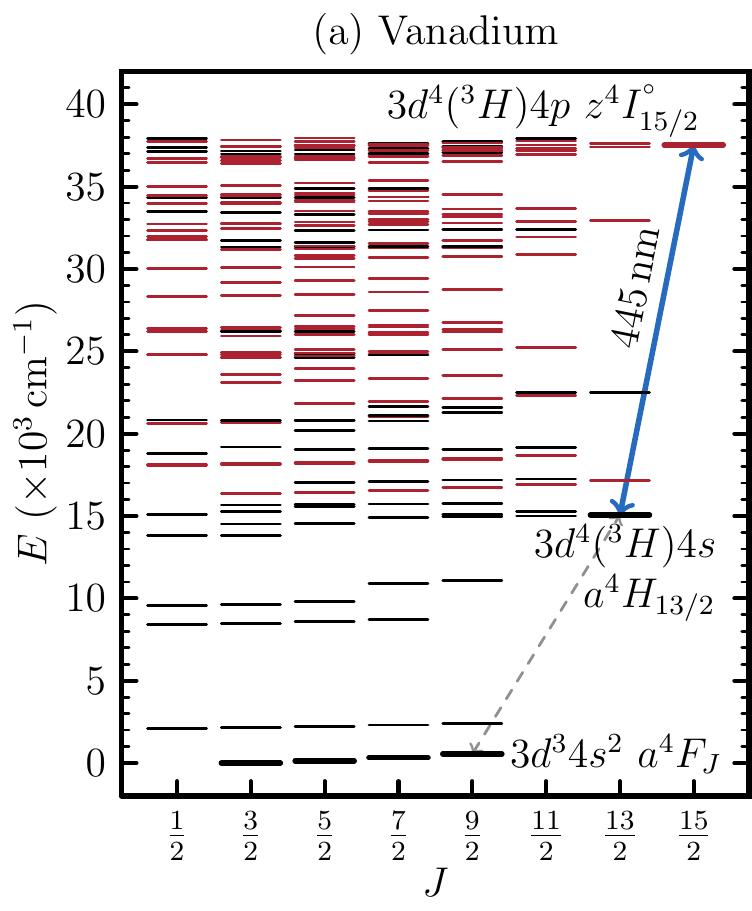}\hspace{0.002\textwidth}\includegraphics[width=0.2189\textwidth]{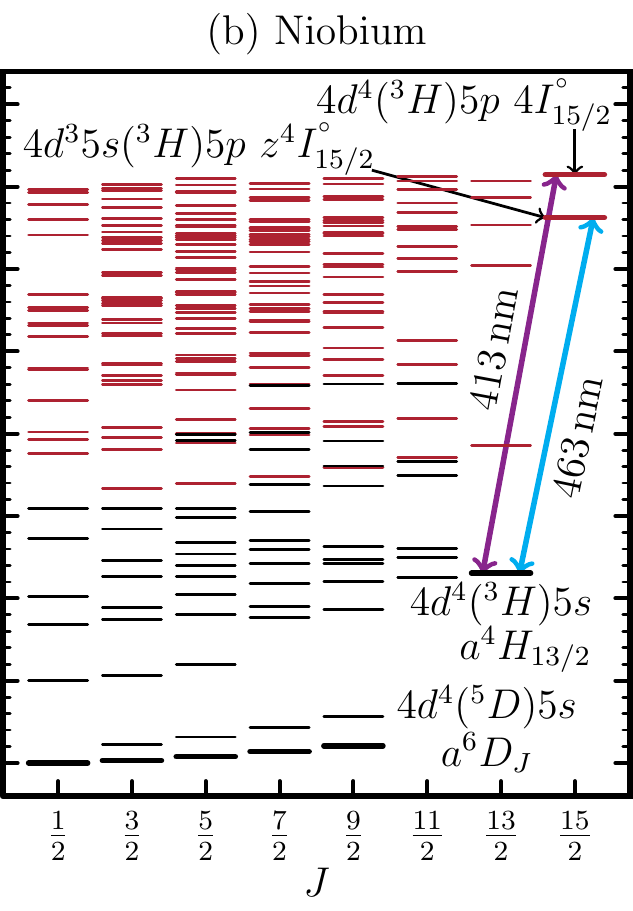}
    \caption{Energy levels of vanadium group elements (a) V and (b) Nb. The fine structure ground states ($a^4F$ for V,  $a^6D$ for Nb), metastable laser-cooling state $a^4H_{13/2}$, and $z^4I^{^\circ}_{15/2}$ state are indicated. The laser-cooling transitions (wavelengths 445 nm for V, 463 nm for Nb) are also indicated. In dashed gray is indicated the dominant decay path of the metastable state of Nb atoms, an E2 transition with $A_{ki}\sim2\pi\times$130 mHz \cite{nist19}}
    \label{fig:V_group_levels}
\end{figure}

In both vanadium (V) and niobium (Nb), the alkali-like, even-parity state with stretched $J$, the $(n-1)d^4(^3H)\, ns^1$ $a^4H_{13/2}$ state, is viable for laser cooling. In both atoms, it connects to cycling transitions with the $(n-1)d^4(^3H)\, np^1$ $z^4I^{^\circ}_{15/2}$, while Nb possesses an additional cooling transition.

\paragraph{Vanadium:} For V, the laser-cooling transition occurs at wavelength $445$ nm with a linewidth of $\Gamma=2\pi\times13$ MHz. There is only one other $J=13/2$ even parity state where leakage from this cycling transition could occur, we estimate the leakage to be $2\times10^{-6}$. The radiative lifetime of the $a^4H_{13/2}$ state has been calculated to be $\sim1.2$ s \cite{nist19}, occurring via electric quadrupole decay to the $a^4F_{9/2}$ state.

\paragraph{Niobium:}For Nb, two laser cooling transitions exist from the comparable metastable state: the laser-cooling transition comparable to the V transition occurs at wavelength $463$ nm between the $a ^4F_{13/2}$ and $z ^4I_{15/2}^{^\circ}$ states. However, there is an additional transition at $413$ nm between the metastable state and the $4d^4( ^3H)5p$ $^4I_{15/2}^{^\circ}$ Neither linewidth is known. However, there are no other $J=13/2$, even parity states in Nb available for leakage. Thus, we expect the cycling transition, once observed, will be fully closed with respect to electric dipole decays. However unobserved even parity states could in principle introduce leakage channels not listed in atomic structure databases \cite{nist19}. In this case, the electric quadrupole decay would be the dominant single particle loss mechanism for trapped atoms.

\subsection{Manganese group}

\begin{figure}
    \centering
    \includegraphics[width=0.26\textwidth]{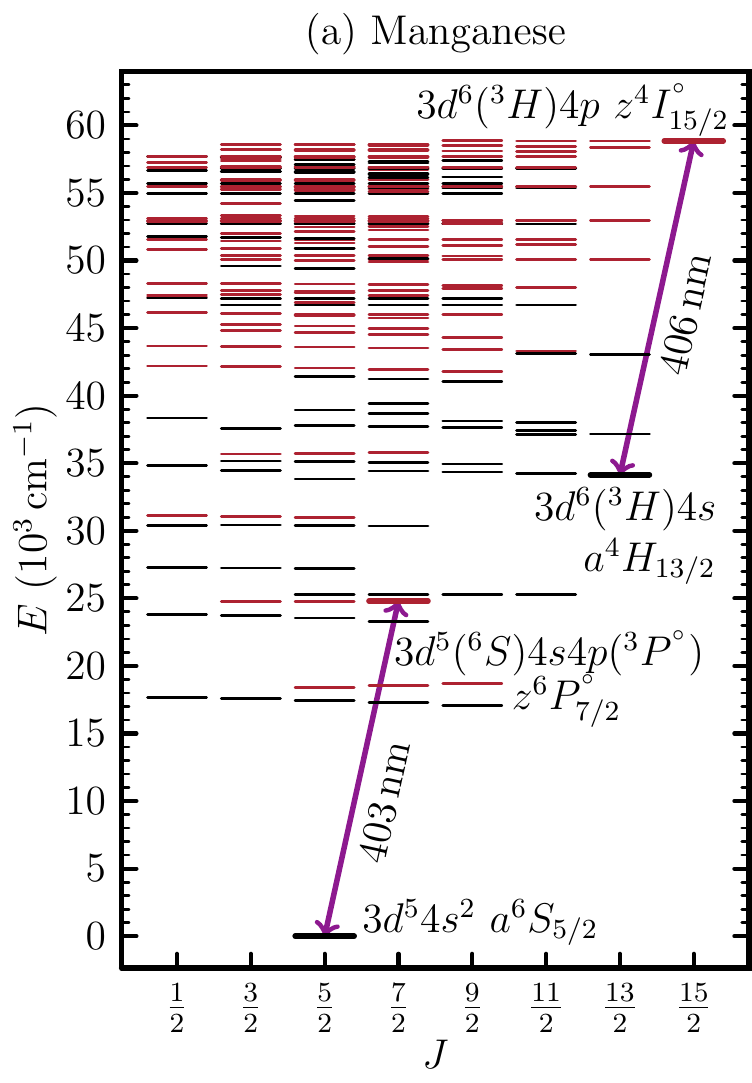}\hspace{0.002\textwidth}\includegraphics[width=0.2189\textwidth]{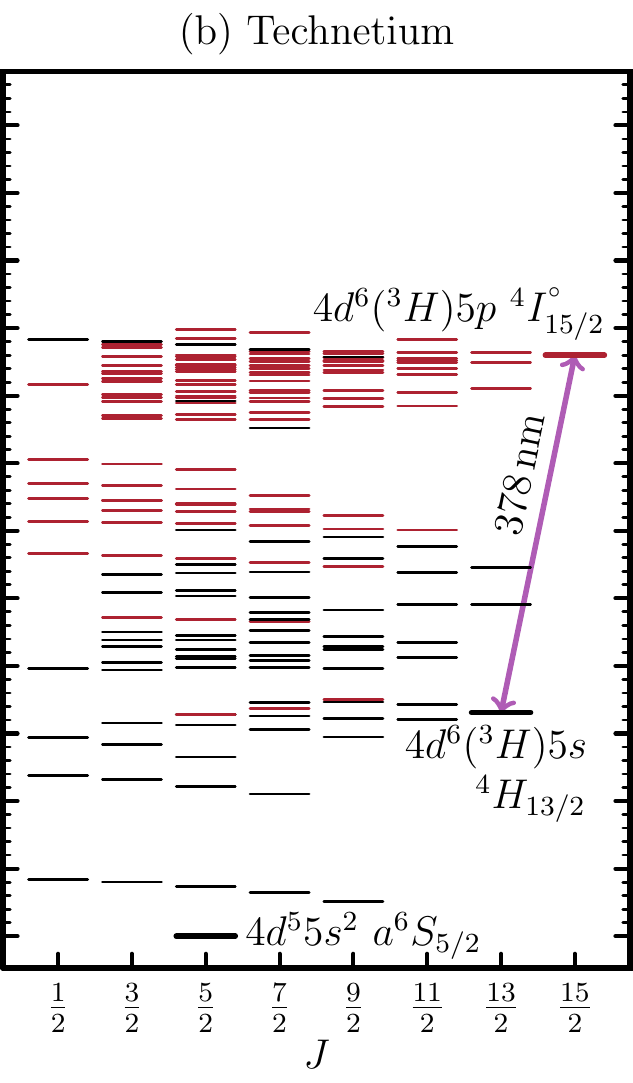}
    \caption{Energy levels of manganese group elements (a) Mn and (b) Tc. Both Mn and Tc possess a viable cycling transition (wavelengths 406 nm and 378 nm) between the metastable $a^4H_{13/2}$ state and the $z^4I_{15/2}^{^\circ}$ state. Mn is also likely laser-coolable directly in the ground state. The three possible laser-cooling transitions are at a wavelength of 403 nm and connect the $a^6S_{5/2}$ ground state to the $z^6P^{^\circ}_J$ states.}
    \label{fig:Mn_group_levels}
\end{figure}

Elements of the manganese group, manganese (Mn) and technetium (Tc), have ground state electronic configurations $(n-1)d^5\, ns^2$ and term $a^6S_{5/2}$ (Fig.\ \ref{fig:Mn_group_levels}). The stretched $J$, alkali-like metastable state potentially suitable for laser cooling is the $(n-1)d^6\, ns$ $a^4H_{13/2}$ state. 

\paragraph{Manganese:} The laser-cooling transition of the metastable state in Mn occurs at a wavelength of $406$ nm.  Its linewidth has not been measured. We expect this transition to be nearly closed, with an estimated leakage of $2\times10^{-5}$. Mn presents a unique challenge for optical pumping due the high energy of the metastable $a^4H_{13/2}$ state with respect to the $3d^5\, 4s^2$ $a^6S_{5/2}$ ground state (34139 cm$^{-1}$).

However, this ground state may be interesting in its own right, as it possesses several nearly closed electronic transitions with wavelengths near 403 nm that connect to the $3d^5(^6S)\, 4s\, 4p(^3P^{^\circ})$ $z^6P^{^\circ}_J$ states. while these states likely have spin-allowed leakage channels, the low frequency of these leakage transitions, and the fact that these leakages involve inner electron transitions, may imply that the branching ratio for leakage is very low.

\paragraph{Technetium:}

Unlike in Mn, there is not a closed transition in the ground state of Tc. Indeed, for Tc, contrarily to Mn, the ground state and lowest excited states are much higher in energy relative to the other even parity metastable states that an atom could leak into. Thus, if one attempted to cycle many photons in the ground state of Tc, it would rapidly leak into a large number of metastable states and be rapidly lost from the trap. However the metastable cooling state ($4d^6(^3H)\, 5s$ $^4H_{13/2}$) is of much lower energy (16553 cm$^{-1}$) than in Mn, and, thus, is likely easier to access. The laser-cooling transition from this state connects to the $4d^6(^3H)\, 5p$ $^4I_{15/2}^{^\circ}$ state and  occurs at the wavelength 378 nm.  Its linewidth has not been reported. Tc is a radioactive element that is not naturally occurring.  However, it does have several long-lived isotopes, including bosonic $^{97}$Tc and fermionic $^{98}$Tc, with lifetimes over $10^6$ years, and also the bosonic $^{99}$Tc - commonly used in medicine - with a lifetime over $10^5$ years.

\section{Conclusion}

By examining the structure of ground and metastable states of transition-metal atoms, we have identified a number of optical transitions that resemble the principal atomic transitions of alkali atoms, in that a single s-orbital valence electron is excited to the p-orbital at the same principal quantum number.  Thus, these transition-metal optical transitions appear suited for laser cooling, i.e.\ they have large oscillator strength, are nearly cycling with little leakage to additional states, and have the right $J\rightarrow J+1$ angular momentum structure to support many standard laser-cooling techniques.  We focus most of our attention on the example of atomic titanium, given that the level structure for this atom is well characterized, that the atom has numerous abundant bosonic and fermionic isotopes, and that its ground state presents attractive features for ultracold atomic physics experiments.  The example set by titanium allows us to sift through the level structures of ten other transition-metal elements, and to identify laser-cooling transitions for each.

There remain many potential pitfalls in applying laser cooling to these atomic elements. One possibility is the loss of atoms due to photoionization by laser cooling light. For five of the elements identified (La, V, Nb, Mn, Tc) the excited state in the laser cooling path is high enough in energy that an additional laser cooling photon can drive the atom over the first ionization limit. While a calculation of photoionization rates of excited states of atoms is beyond the scope of this paper, such photoionization could present a mechanism to limit the lifetime of atoms in the MOT in addition to the leakage out of the laser cooling transition and the decay of the metastable state.

Another potential pitfall is that the atoms in metastable states may be subject to strong collisional decay, including light-induced decay in the presence of laser-cooling and optical-pumping light.  Even if laser cooling is successful, further cooling toward quantum degeneracy may require multi-step optical pumping to the electronic ground state.  For some elements, inelastic decay channels in the ground state may be unexpectedly high.  Nevertheless, with such a large selection of elements (11) and of stable (26 total) and long-lived (9 total with $>10^6$ yr half-life) isotopes potentially made available for ultracold atomic physics experiments, the investment of efforts to realize laser-cooling of these elements is warranted.



We are grateful to M. Lepers et al.\ \cite{lepe14er} for allowing us to use their data on erbium transitions and also to D.\ Pe\~na, A.\ Neely, and M.\ Aguirre for discussions and experimental insights.  We acknowledge support from the Heising-Simons Foundation, from the ARO-STIR program (Contract No.\ W911NF1910017), and from the California Institute for Quantum Entanglement supported by the Multicampus Research Programs and Initiative of the UC Office of the President (Grant No.\ MRP-19-601445).


\bibliography{allrefs_x2}

\newpage

\end{document}